\documentclass{revtex4}
\usepackage{graphicx}
\usepackage{fancyhdr}
\usepackage{amsmath}
\usepackage{color}

\begin{document}

\title{Gamma-rays Constraint on Higgs Production from Dark Matter
Annihilation}

%

\author{Takashi Toma}
\affiliation{
Institute for Particle Physics Phenomenology University of Durham,
Durham DH1 3LE, UK
}

\begin{abstract}
A gamma line signal in cosmic-ray observation would be found 
 when Higgs is produced at rest or almost rest by DM annihilation
 because of decay of $126~\mathrm{GeV}$ Higgs into two gamma. This line signal
 gives a constraint on Higgs production cross section by DM. We
 examine this point with simple analysis in this work. 
\end{abstract}

\maketitle

\thispagestyle{fancy}


\section{Introduction}
The Standard Model (SM)-like Higgs boson was discovered at LHC and the 
mass is $m_H\approx126~\mathrm{GeV}$~\cite{Chatrchyan:2012twa,
Chatrchyan:2012tx, ATLAS:2012ad, ATLAS:2012ae, Chatrchyan:2012ufa}. 
This result shows good agreement with the SM, however we do not know yet
whether it is truly Higgs in the SM or not. Consequently more information is
required in order to define it. One of possibilities to gain more 
knowledge about Higgs is examining interaction between Higgs and Dark
Matter (DM). This is achieved with astrophysical data such as cosmic-rays. 
The discovered Higgs at LHC decays into gamma and it may be observed by
such as Fermi-LAT if Higgs is produced by DM
annihilation~\cite{Bringmann:2012vr, Weniger:2012tx}. Especially, 
if Higgs at rest or almost rest is produced by DM, a
line signal at $E_{\gamma}\approx63~\mathrm{GeV}$ is emitted due to the
Higgs decay channel $H\to2\gamma$ which can be characteristic signal in
gamma-ray observations. 
According to comparison with observed gamma-ray data, we can obtain 
constraint on annihilation cross section into Higgs, and finally
coupling strength between DM and Higgs would be constrained. 
In this talk, we focus on rest or slightly boosted Higgs production by
DM annihilation~\cite{Bernal:2012cd}. 

\section{gamma-ray from DM annihilation}
There are two types of $\gamma$-rays. The first one is $\gamma$-ray of
continuum energy spectrum which comes from Final State Radiation (FSR), decays and
hadronization of the final state particles. The second one is
$\gamma$-ray line spectrum which is directly produced and
decay of rest Higgs generated by DM annihilation. 
In the case of directly produced $\gamma$-ray, the energy is fixed to
$E_{\gamma}=m_{\chi}$ for the process $\chi\chi\to\gamma\gamma$ where
$\chi$ is DM and such a process is always loop-suppressed. 
The energy of gamma generated by decay of rest Higgs is almost fixed
to $E_{\gamma}=63~\mathrm{GeV}$. Namely, we can investigate whether a line
signal around $63~\mathrm{GeV}$ via rest Higgs decay exists or not by
comparing with data. Then we obtain a 
constraint on annihilation cross section into Higgs when DM mass is some
specific values we will discuss later. 

To compare line and continuum $\gamma$-rays, we have to take into
account phase space suppression. For example, for
the process $\chi\chi\to HH$, the mass relation $m_{\chi}\approx m_{H}$
is required to produce rest Higgs. Hence phase space
suppression factor $\sqrt{1-m_{H}^2/m_{\chi}^2}$ should be multiplied. On the
other hand, we do not have that factor for the process
like $\chi\chi\to b\bar{b}$ since the bottom quark mass is light enough
compared with DM mass. As a result much broad $\gamma$ spectrum is
expected via the $b\bar{b}$ channel. In addition, the ratio of branching
fractions among them is small as follows
\begin{equation}
\frac{\mathrm{Br}\left(H\to\gamma\gamma\right)}
{\mathrm{Br}\left(H\to b\bar{b}\right)}
\approx4\times10^{-4}.
\end{equation}
Therefore the $\gamma$-ray line from the Higgs decay might be swamped by
the continuum $\gamma$ spectrum unless some enhancement mechanism is
taken into account. 
In the case of SUSY models for example, a way-out is making use of a
resonance of pseudo-scalar ($\phi_A$) mediation in s-channel process like
$\chi\chi\to \phi_A\to HH$. Moreover $b\bar{b}$
production should be suppressed by heavy sbottom and small coupling
between bottom and the pseudo-scalar. As a result it is possible to see a
line signal from the rest Higgs decay. 

From now on, we focus on some specific DM masses such as
$63~\mathrm{GeV}$, $109~\mathrm{GeV}$ and $126~\mathrm{GeV}$ to generate
Higgs at rest. 
In the case of $m_{\chi}=63~\mathrm{GeV}$, the target channel is
$\chi\chi\to H\gamma$ which is possible when DM spin is $1/2$ or
$1$. This fact can be seen from spin statistics and no longitudinal component
of photon. 
After the production of Higgs, The Higgs decays into
$\gamma\gamma$ and the energy is $63~\mathrm{GeV}$. 
In addition, the energy of the direct produced $\gamma$ is
$E_{\gamma}=(4m_{\chi}^2-m_{H}^2)/4m_{\chi}$ and this is expected to be
too low energy to detect it. 
If $\gamma$ excess is found at $63~\mathrm{GeV}$ in experiments, it
might be confused with direct produced $\gamma$ like $\chi\chi\to\gamma\gamma$
when DM mass is $m_{\chi}=126~\mathrm{GeV}$. Therefore we have to disentangle it to
identify where the observed signal comes from. 
For $109~\mathrm{GeV}$ DM, the target channel is $\chi\chi\to HZ$. A
peak is expected at $63~\mathrm{GeV}$ coming from the rest Higgs decay. 
For example, this process can be possible via t-channel and s-channel
exchange diagrams if DM interacts with $Z$ boson directly. 
In addition to the Higgs decay signal, a peak at $109~\mathrm{GeV}$ is
expected via direct $\gamma$ production by DM annihilation
$\chi\chi\to\gamma\gamma$. Similarly, $72~\mathrm{GeV}$ line comes from
$\chi\chi\to\gamma{Z}$. However they are not relative with Higgs, so we
do not focus on them here.
In the case of $126~\mathrm{GeV}$ DM, the channel $\chi\chi\to HH$ is
possible to generate the rest Higgs. This
channel also gives $63~\mathrm{GeV}$ line signal. In SUSY models, this process can
be possible via box diagram of bottom and sbottom at one-loop, moreover 
pseudo-scalar s-channel exchange at tree level if DM has interaction
with pseudo-scalar. As well as the previous case, we have to disentangle
with direct $\gamma$ production like $\chi\chi\to\gamma\gamma,\:H\gamma$
when DM mass is $63~\mathrm{GeV}$ when such a line is experimentally found. 

\begin{figure}[t]
\begin{center}
\includegraphics[width=50mm]{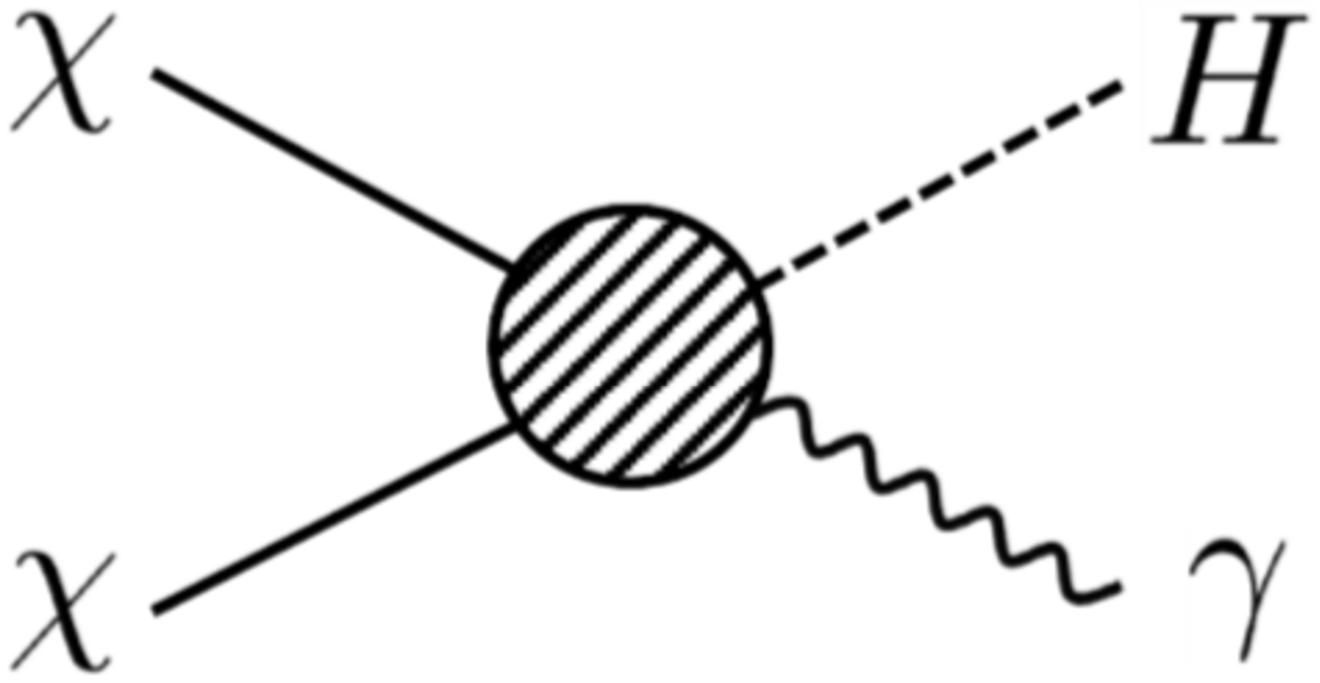}
\hspace{0.5cm}
\includegraphics[width=50mm]{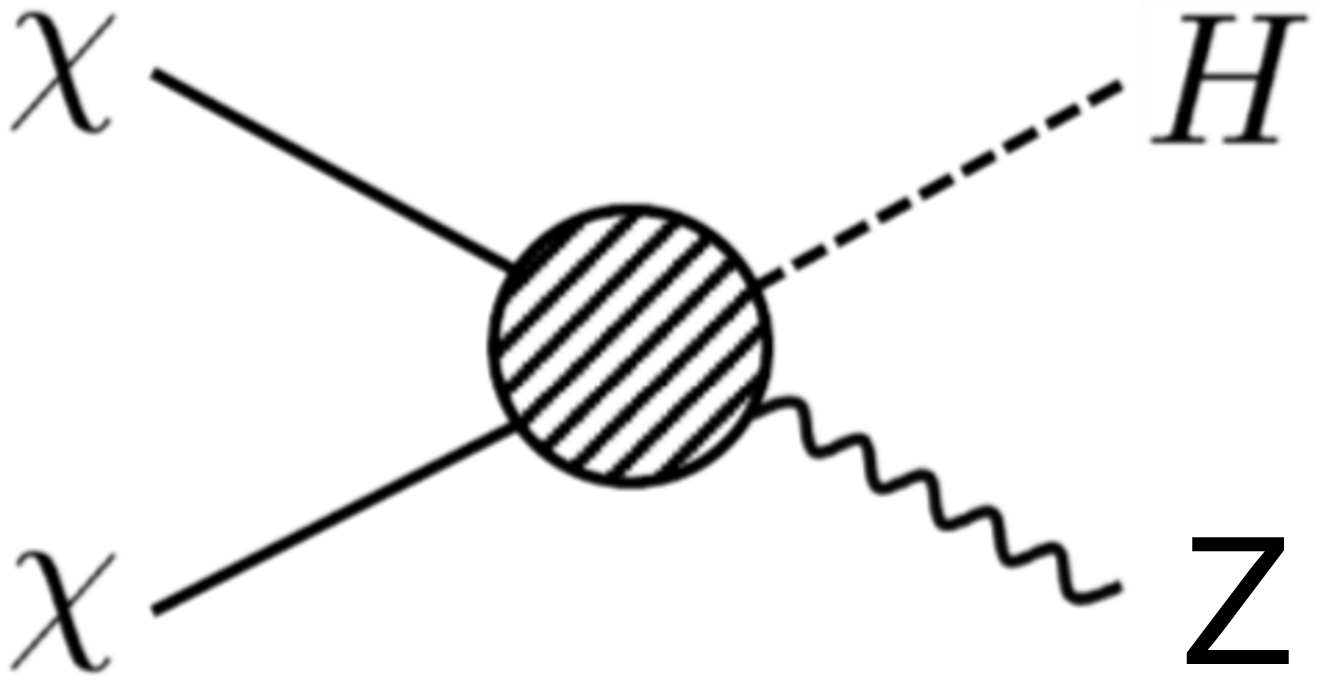}
\hspace{0.5cm}
\includegraphics[width=50mm]{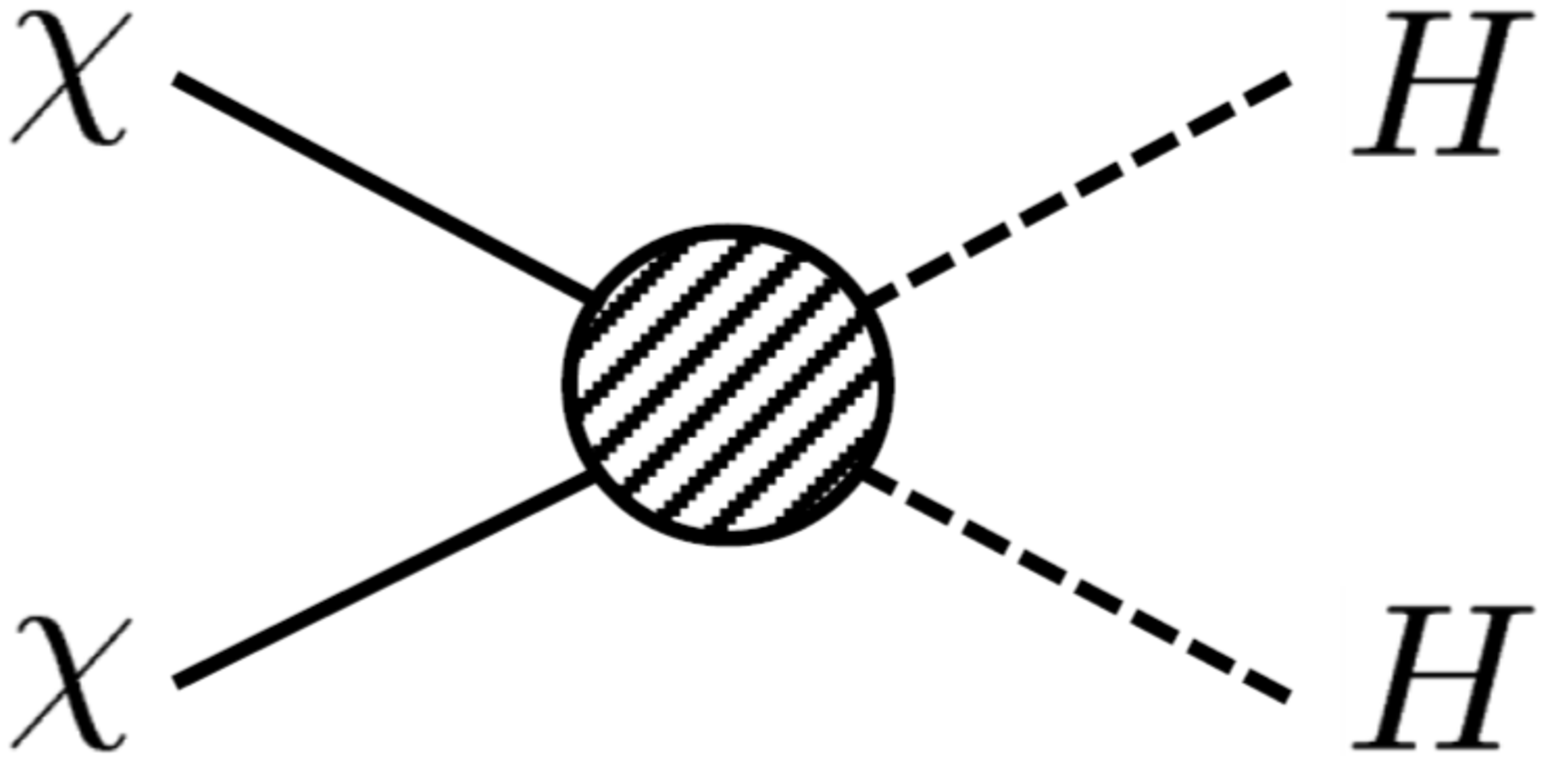}
\caption{The DM annihilation channels with the rest Higgs in final state
 that we consider in this work.}
\label{fig:diagram}
\end{center}
\end{figure}

\section{gamma-ray flux}
The $\gamma$-ray flux coming from DM annihilation is given as 
\begin{equation}
\frac{d\Phi_\gamma}{dE_\gamma}=
\eta\frac{\left<\sigma{v}\right>}{8\pi m_{\chi}^2}\frac{dN_\gamma}{dE_{\gamma}}
\int_{\Delta\Omega}d\Omega\int_{\mathrm{los}}
\rho^2\left(r\left(s,\Omega\right)\right)ds.
\label{eq:flux}
\end{equation}
where $\eta$ is symmetry factor which is $1$ for self-conjugate DM and
$1/2$ for non self-conjugate DM, $\left<\sigma{v}\right>$ is DM
annihilation cross section into $\gamma$, $dN_{\gamma}/dE_{\gamma}$ is energy
spectrum and $\rho(r)$ is DM profile which is taken as Einasto
profile~\cite{einasto} in this analysis, and the profile is expressed as 
\begin{equation}
\rho(r)=0.193\rho_{\odot}\mathrm{exp}
\left[-\frac{2}{\alpha}\left(\left(\frac{r}{r_s}\right)^{\alpha}-1\right)
\right],
\end{equation}
where $\alpha=017$, $r_s=20~\mathrm{kpc}$,
$\rho_{\odot}=0.386~\mathrm{GeV/cm^3}$. 
Note that the energy dependence comes from only the energy spectrum
$dN_{\gamma}/dE_{\gamma}$. 

\subsection{$\gamma$-ray from Higgs at rest}
\begin{figure}[t]
\begin{center}
\includegraphics[width=60mm]{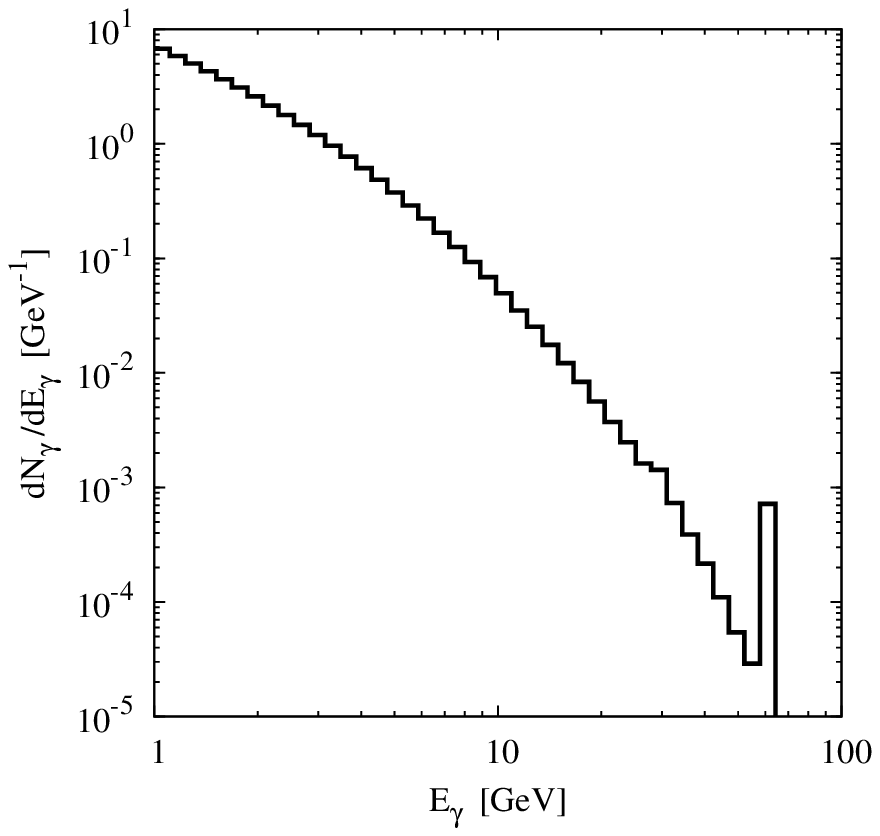}
\hspace{1cm}
\includegraphics[width=78mm]{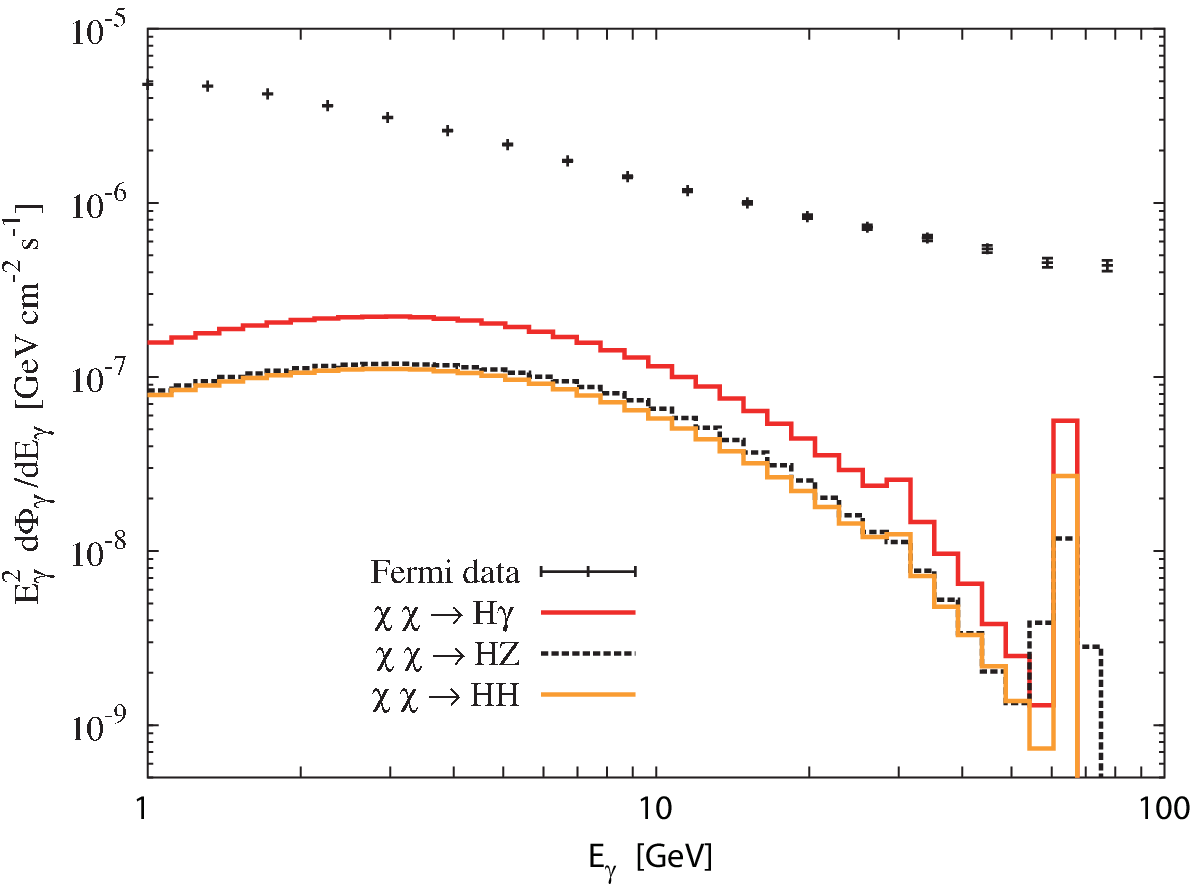}
\caption{The energy spectrum of $\gamma$ coming from the decay of the
 rest Higgs (left) and the $\gamma$-ray flux calculated by using the energy
 spectrum for different three channels $\chi\chi\to H\gamma,\:HZ,\:HH$ (right).}
\label{fig:rest}
\end{center}
\end{figure}

We calculate the energy spectrum of $\gamma$ via the rest Higgs is
calculated by PYTHIA 6.4~\cite{Sjostrand:2006za} and that is shown in the left hand side of
Fig~\ref{fig:rest}. 
All channels of Higgs decay such as $H\to b\bar{b},\:WW^*,\:ZZ^*$ etc are included in
the figure. 
A peak at $63~\mathrm{GeV}$ can be seen and the peak comes from the
Higgs decay process $H\to2\gamma$. A small bump around $30~\mathrm{GeV}$
which is due to the Higgs decay $H\to Z\gamma$ is found, and the energy is expressed as
$E_{\gamma}=\left(m_H^2-m_Z^2\right)/2m_H$. 
The $\gamma$ line signal is visible despite of the small branching
fraction of $H\to\gamma\gamma$. 

The $\gamma$-ray flux from Higgs decay is shown in
the right hand side of Fig.~\ref{fig:rest} with Fermi-LAT data. Three channels $\chi\chi\to
H\gamma,\:HZ,\:HH$ are included in  the figure. The most visible signal is
obtained for the channel $\chi\chi\to H\gamma$. This is
because DM mass is fixed to specific values for each channel and the
mass is smallest for this channel, moreover the flux is proportional to
$m_{\chi}^{-2}$ as can be seen in Eq.~(\ref{eq:flux}). 
We can obtain the limits for DM annihilation into Higgs as
\begin{eqnarray}
\left<\sigma{v}\right>&\lesssim&
2.5\times10^{-25}~\mathrm{cm^3/s}\quad\mathrm{for}\quad \chi\chi\to
H\gamma,\quad m_\chi=63~\mathrm{GeV},\\
\left<\sigma{v}\right>&\lesssim&
5.0\times10^{-25}~\mathrm{cm^3/s}\quad\mathrm{for}\quad \chi\chi\to HZ,
\quad m_\chi=109~\mathrm{GeV},\\
\left<\sigma{v}\right>&\lesssim&
6.0\times10^{-25}~\mathrm{cm^3/s}\quad\mathrm{for}\quad \chi\chi\to HH,
\quad m_\chi=126~\mathrm{GeV}.
\end{eqnarray}
These limits correspond to roughly $10^{-28}\sim10^{-27}\mathrm{cm^3/s}$
for $\gamma$ production by DM annihilation.
The limits are obtained from the line excess of $H\to 2\gamma$ rather
than the other continuum $\gamma$. 

\begin{figure}[t]
\begin{center}
\includegraphics[width=60mm]{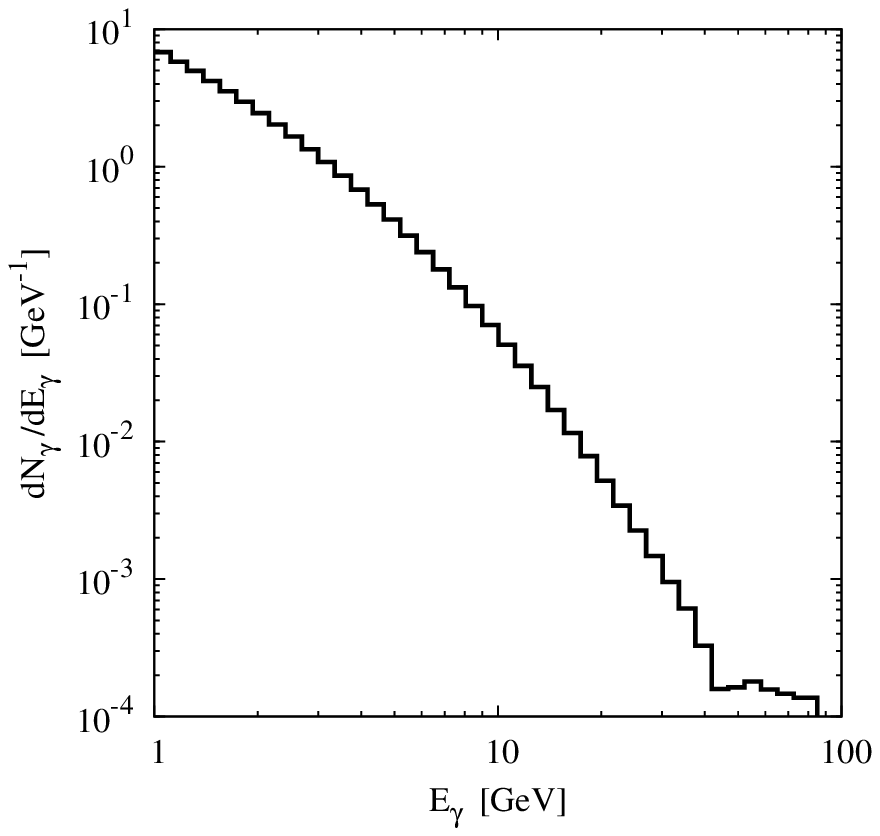}
\hspace{1cm}
\includegraphics[width=78mm]{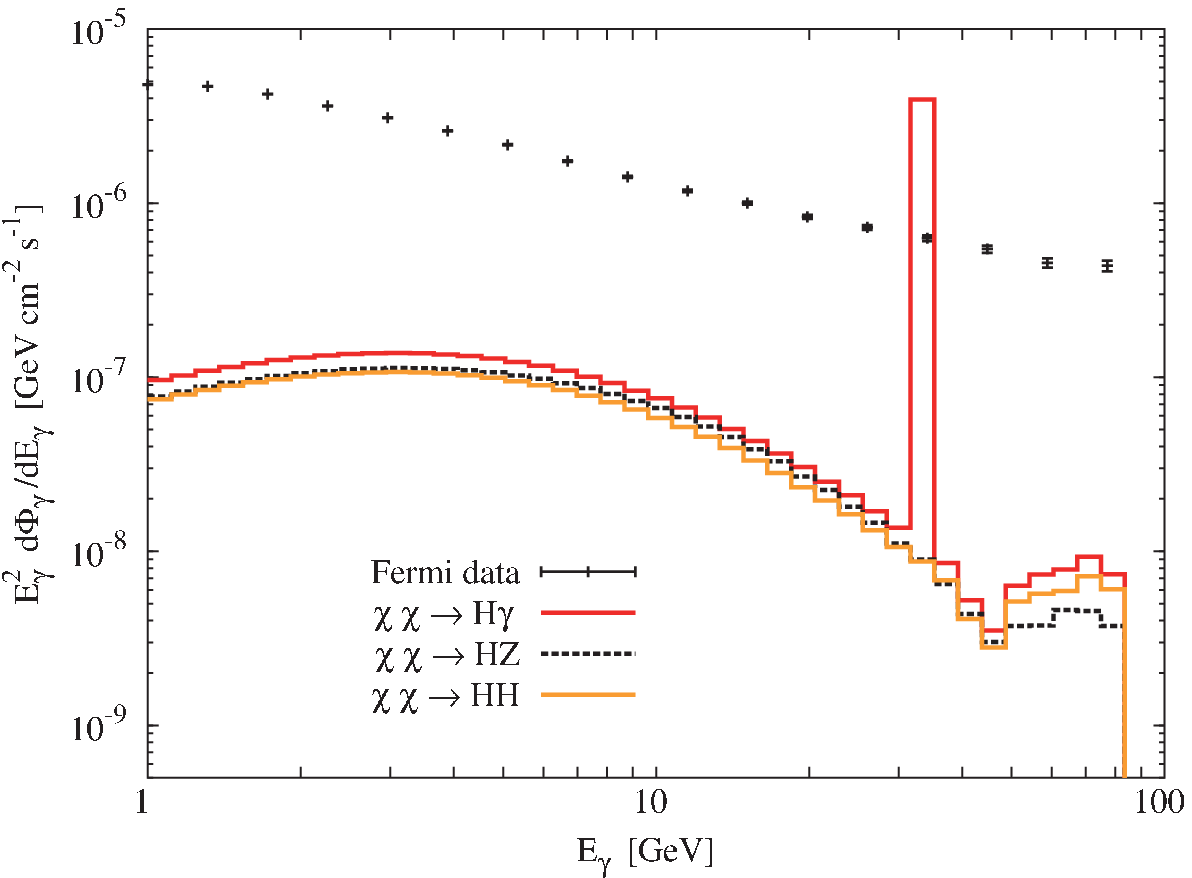}
\caption{The energy spectrum of $\gamma$ coming from the decay of the
 slightly boosted Higgs (left) and the $\gamma$-ray flux calculated by using the energy
 spectrum for different three channels $\chi\chi\to H\gamma,\:HZ,\:HH$
 (right). Here Higgs is boosted to $E_H=130~\mathrm{GeV}$. The DM mass
 is fixed to $m_{\chi}=81,\:178,\:130~\mathrm{GeV}$ for the channels $\chi\chi\to
 H\gamma,\:HZ.\:HH$ respectively.}
\label{fig:boost}
\end{center}
\end{figure}

\subsection{$\gamma$-ray from slightly boosted Higgs}
The energy spectrum from boosted Higgs is shown in the left hand side of Fig~\ref{fig:boost}. 
The peak around $63~\mathrm{GeV}$ via $H\to\gamma\gamma$ becomes tiny
because of the broad energy distribution of $H\to\gamma\gamma$ in the
boosted case. No small
excess around $30~\mathrm{GeV}$ which is
different from the rest case due to the same reason. The energy spectrum
of $H\to2\gamma$ is smaller than the other continuum spectrum with the
several order of magnitude.

The $\gamma$-ray flux is depicted in the right hand side of
Fig~\ref{fig:boost}. We have an intense line signal at $32~\mathrm{GeV}$ for
$\chi\chi\to H\gamma$. This line gives the limit of
$\left<\sigma{v}\right>\lesssim4\times10^{-27}~\mathrm{cm^3/s}$ for the
channel $\chi\chi\to H\gamma$. 
This strong line comes from direct $\gamma$ production from DM
annihilation. The direct $\gamma$ energy appears at
$E_{\gamma}=(4m_{\chi}^2-m_H^2)/4m_\chi$. We can see from the equation
that no such a strong line for the rest Higgs,
but we have the line for boosted case. 
For the other two channels $\chi\chi\to HZ,\:HH$, we can get the rough limits of
$\left<\sigma{v}\right>\lesssim5\times10^{-25}~\mathrm{cm^3/s}$. 

\section{Summary}
The SM-like Higgs boson was discovered at LHC. Using the decay property
of the SM-like Higgs, we obtained the constraint on Higgs production
cross section by DM in cases of some DM masses. The constraint is
roughly $\left<\sigma{v}\right>\lesssim10^{-25}~\mathrm{cm^3/s}$ for all
the channels when Higgs is generated at rest or slightly boosted frame. 
Especially, the limit will be important when an enhancement mechanism of production cross
section of Higgs such as Sommerfeld enhancement or resonance enhancement is working. 
We have done only rough estimation in this work. More precious constraint
will be obtained by more detail analysis. 

\begin{acknowledgments}
This work is supported from the European ITN project
 (FP7-PEOPLE-2011-ITN, PITN-GA-2011-289442-INVISIBLES). 
The numerical calculations were carried out on SR16000 at YITP
in Kyoto University.
\end{acknowledgments}

\bigskip 

\begin{thebibliography}{99} 
\bibitem{Chatrchyan:2012twa} 
  S.~Chatrchyan {\it et al.}  [CMS Collaboration],
  Phys.\ Lett.\ B {\bf 710}, 403 (2012)
  [arXiv:1202.1487 [hep-ex]].

\bibitem{Chatrchyan:2012tx} 
  S.~Chatrchyan {\it et al.}  [CMS Collaboration],
  Phys.\ Lett.\ B {\bf 710}, 26 (2012)
  [arXiv:1202.1488 [hep-ex]].

\bibitem{ATLAS:2012ad} 
  G.~Aad {\it et al.}  [ATLAS Collaboration],
  Phys.\ Rev.\ Lett.\  {\bf 108}, 111803 (2012)
  [arXiv:1202.1414 [hep-ex]].

\bibitem{ATLAS:2012ae} 
  G.~Aad {\it et al.}  [ATLAS Collaboration],
  Phys.\ Lett.\ B {\bf 710}, 49 (2012)
  [arXiv:1202.1408 [hep-ex]].

\bibitem{Chatrchyan:2012ufa} 
  S.~Chatrchyan {\it et al.}  [CMS Collaboration],
  Phys.\ Lett.\ B {\bf 716}, 30 (2012)
  [arXiv:1207.7235 [hep-ex]].

\bibitem{Bringmann:2012vr} 
  T.~Bringmann, X.~Huang, A.~Ibarra, S.~Vogl and C.~Weniger,
  JCAP {\bf 1207}, 054 (2012)
  [arXiv:1203.1312 [hep-ph]].

\bibitem{Weniger:2012tx} 
  C.~Weniger,
  JCAP {\bf 1208}, 007 (2012)
  [arXiv:1204.2797 [hep-ph]].

\bibitem{Bernal:2012cd} 
  N.~Bernal, C.~Boehm, S.~Palomares-Ruiz, J.~Silk and T.~Toma,
  arXiv:1211.2639 [hep-ph].

\bibitem{einasto}
J. Einasto, Trudy Inst. Astrofiz. Alma-Ata 51, 87 (1965).

\bibitem{Sjostrand:2006za} 
  T.~Sjostrand, S.~Mrenna and P.~Z.~Skands,
  JHEP {\bf 0605}, 026 (2006)
  [hep-ph/0603175].

\end{thebibliography}

\end{document}